\documentstyle[12pt,aasms4]{article}

\begin{document}

\title{Stokes Parameters in\\ Cosmic Microwave Background Measurements}
\author{Ka Lok Ng\altaffilmark{1} and Kin-Wang Ng\altaffilmark{2}}
\affil{Institute of Physics,
Academia Sinica, Taipei, Taiwan 11529, R.O.C.}

\altaffiltext{1}{Present address: Silicon Graphics Limited, Taipei, Taiwan, R.O.C.\\
Email: albert@taiwan.sgi.com}
\altaffiltext{2}{Email: nkw@phys.sinica.edu.tw}

\begin{abstract}
We compute numerically the scalar- and tensor-mode induced Stokes parameters of the cosmic microwave background, by taking into account the basis rotation effects. It is found that the tensor contribution to the polarization power spectrum get enhanced and dominates over the scalar contribution for low multipoles in a universe with or without recombination. Furthermore, we show that all full-sky averaged two-point cross-correlation functions of the Stokes parameters vanish. We thus comment on the cross-correlation between the anisotropy and polarization of the CMB, and calculate the expected signal to noise ratio for the polarization experiment underway.
\end{abstract}


\section{Introduction}

The detection of the large-angle anisotropy of the cosmic microwave
background by the {\it COBE} DMR (\cite{smo92}) is an important evidence for the large-scale space-time inhomogenities. Since then, several small-scale
anisotropy measurements have hinted that the Doppler
peak resulted from acoustic oscillations of the baryon-photon plasma on the
last scattering surface seems to be present. It is believed that measuring CMB
anisotropies on all angular scales can give invaluable information of the early universe, such as testing inflation, pinning down
cosmological parameters, and differentiating scalar from tensor perturbations
(see, e.g., \cite{ste94}).

Complementary to the anisotropy,
polarization is another important property of the CMB.
The degree of CMB polarization has been extensively calculated (see \cite{ng96} and references therein). 
In a universe with standard recombination, the
polarization is about $10\%$ of the anisotropy on small angular scales while
the large-scale polarization is insignificant. If early matter reionization of
the universe occurs, the large-scale polarization would be greatly
enhanced to a $10\%$ level with the large-scale anisotropy reasonably
unaffected. Meanwhile, the small-scale anisotropy would be significantly
suppressed. Thus, measuring CMB polarization might provide additional
information about the early universe. So far, only experimental upper limits
have been set (\cite{lub83}; \cite{par88}; \cite{wol93}), and
the signal level is expected to be achieved through the use of new technology
and instrument design (P. T. Timbie 1995, private communication).

In previous calculations of the degree of CMB polarization, approximations have more or less been made. For instances,
the separation angle in the two-point auto-correlation
functions has been assumed small, or/and the basis rotation
from the ${\bf\hat k}$-basis of the perturbation to the fixed basis of an
experiment has been totally neglected.
Although it is expected that these approximations would
be valid for small-angle polarization calculations, they may break down on
large-scale calculations. In this {\it Letter}, we will compute the Stokes parameters by taking into account the basis rotation effects. Also, we will consider the two-point cross-correlation functions of the Stokes parameters.

\section{Stokes Parameters of the CMB}

Polarized light is conventionally described in terms of four Stokes parameters: $(I,Q,U,V)$, each of which is a function of the observational position in the universe, $\bf r$, as well as a unit vector pointing to the celestial sphere, $\bf \hat e$.
In polarization measurements, it is convenient to choose the North
Celestial Pole as the reference axis, $\bf \hat z$, and the linear polarization of the CMB is  measured in terms of $Q$ and $U$ defined by:
$Q=T_{N,S}-T_{E,W}$ and $U=T_{NE,SW}-T_{NW,SE}$,
where $T_{N,S}$ is the antenna temperature of radiation polarized along the
north-south direction and so on (\cite{lub83}). Let us
define $T=I-{\bar I}$ as the temperature fluctuation about the mean. It is 
known that $V$ would decouple from the other components. Thus, we will
calculate the Stokes components $(T,Q,U)$ of the CMB induced by scalar
and tensor perturbations. We first decompose $(T,Q,U)$
into Fourier components $(T_{\bf k},Q_{\bf k},U_{\bf k})$, which are
induced by a Fourier mode with wavevector ${\bf k}$. Then, for a fixed $\bf k$,
we solve the radiative transfer equation for the Stokes
parameters $(T'_{\bf k},Q'_{\bf k},U'_{\bf k})$, which are defined with respect to the $\bf \hat k$ axis. The primed and unprimed parameters are related by a simple rotation (\cite{bon87}),
\begin{equation}
\left ( \begin{array}{c} Q_{\bf k}\\U_{\bf k}  \end{array} \right)=
\left( \begin{array}{cc}
\cos2\psi&\sin2\psi\\
-\sin2\psi&\cos2\psi
\end{array} \right)
\left ( \begin{array}{c}
Q'_{\bf k}\\U'_{\bf k}  \end{array} \right),
\end{equation}
and $T_{\bf k}=T'_{\bf k}$, where the rotation angle $\psi$ is given by
\begin{equation}
\cos\psi=
{{ {\bf \hat k\cdot \hat z}-({\bf \hat k \cdot \hat e})({\bf \hat z \cdot
\hat e}) } \over { \sqrt{1-({\bf \hat k\cdot \hat e})^2} \sqrt{1-({\bf
\hat z\cdot \hat e})^2}}}, \quad
\sin\psi=
{ { {\bf \hat e \cdot }({\bf \hat z \times \hat k}) }
\over { \sqrt{1-({\bf \hat k\cdot \hat e})^2} \sqrt{1-({\bf \hat z\cdot
\hat e})^2} }}.
\end{equation}
In the tensor-mode case, $T'_{\bf k}=(1/2)\alpha(1-\mu^2)\cos2\phi$,
$Q'_{\bf k}=(1/2)\beta(1+\mu^2)\cos2\phi$,
and $U'_{\bf k}=\beta\mu\sin2\phi$
for the $+$ mode solution, where ($\mu=\cos\theta,\phi$) are the polar angles
of the unit vector ${\bf \hat e}$ in a coordinate system with $\bf \hat k$ as
the polar axis. $\alpha$ and $\beta$ are functions of $\mu$ only, usually expanded in terms of Legendre polynomials:
$\alpha(\mu) = \sum _{l} (2l+1) \alpha_l P_l(\mu)$ and 
$\beta(\mu)=\sum _{l} (2l+1) \beta_l P_l(\mu)$,
where $\alpha_l$ and $\beta_l$ have previously been calculated
(see \cite{ng96} and references therein).
The $\times$ mode solution is given by the same expressions except replacing
$\cos2\phi$ by $\sin2\phi$, and $\sin2\phi$ by $-\cos2\phi$.
In the scalar-mode case, $T'_{\bf k}=\alpha$, $Q'_{\bf k}=\beta$,
and $U'_{\bf k}=0$.

\section{Anisotropy and Polarization Power Spectra}

In CMB measurements, the measured Stokes parameters are usually expanded in
terms of the spherical harmonics,
\begin{equation}
T({\bf r},{\bf \hat e})=\sum_{l,m} t_{lm}({\bf r}) Y_{lm}({\bf \hat e}),\quad 
Q({\bf r},{\bf \hat e})=\sum_{l,m} q_{lm}({\bf r}) Y_{lm}({\bf \hat e}),\quad 
U({\bf r},{\bf \hat e})=\sum_{l,m} u_{lm}({\bf r}) Y_{lm}({\bf \hat e}).
\label{tqu}
\end{equation}
From this, we define the anisotropy power spectrum, $C_l\equiv T_l$, and the
polarization power spectrum, $C^P_l\equiv Q_l+U_l$, where
\begin{equation}
T_l\equiv \langle\sum_m t^*_{lm} t_{lm} \rangle = \frac{2l+1}{4\pi}
\int d{\bf\hat e_1}d{\bf\hat e_2} P_l({\bf \hat e_1\cdot \hat e_2})
\langle T_1 T_2 \rangle, 
\label{ps}
\end{equation}
and $Q_l$ and $U_l$ are given by similar expressions.
In equation~(\ref{ps}), the angular brackets denote taking an average over all
observational positions, i.e., $\langle f({\bf r})\rangle \equiv 1/(2\pi)^3
\int d{\bf r} f({\bf r})$, and $T_1$ represents $T({\bf r},{\bf \hat e_1})$.

In the following, we will present the results for the anisotropy power spectrum, $C_l={C_l}^{(S)}+{C_l}^{(T)}$, and the polarization power spectrum,
$C^P_l={C^P_l}^{(S)}+{C^P_l}^{(T)}$, where $S$ and $T$ represent respectively
the scalar and tensor contributions. Specifically, we work in a flat CDM model
with no recombination, in which the Hubble parameter $h=0.5$ and the baryon
density $\Omega_B h^2=0.0125$. Also, we assume that ${C_2}^{(S)}={C_2}^{(T)}$,
the scalar power index is $n_s=0.85$, and the tensor's is $n_t=-0.15$.

To evaluate the power spectra in equation~(\ref{ps}),
we have used the Boltzmann code (\cite{ng96}) to solve
for $\alpha_l$ and $\beta_l$. For the anisotropy, the integral $T_l$
can be easily evaluated and analytic solutions for ${C_l}^{(S)}$ and
${C_l}^{(T)}$ can be found in terms of $\alpha_l$'s (\cite{cri93a}).
For the polarization, unfortunately, the multi-dimensional integrations
involved in $Q_l$ and $U_l$ are difficult.
However, it has been attempted to calculate $Q_l$ and $U_l$ approximately.
In both scalar and tensor cases, by neglecting the basis rotation, i.e.,
setting $\psi=0$, analytic expressions for ${C^P_l}^{(S)}$ and ${C^P_l}^{(T)}$ have been obtained by \cite{ng95}. Using these approximate formulae, we have plotted in Figure~1 the scalar contribution, ${C^P_l}^{(S)}/C_l$,
the tensor contribution, ${C^P_l}^{(T)}/C_l$, and the total polarization.

To study the effect of the basis rotation, we numerically compute the
integrals $Q_l$ and $U_l$ using the Gaussian quadratures integration method. We have computed the power spectra for $l=2,3,5,10,20,30$ with a $10\%$ accuracy (improving the accuracy by an order of magnitude would require much more computer time). We have found that the coordinate-dependent Stokes parameters $Q_l$ and $U_l$ are equal for a fixed $l$, as expected for Gaussian random perturbations. For the scalar mode, including the rotation angle does not apparently change the results from
using the approximate formula with $\psi=0$. On the contrary, in the tensor
case, the multipoles for $l<30$ are modified, and the above approximation is
valid only for $l>30$. This is shown in Figure~1 with solid symbols. Note that
${C^P_2}^{(T)}$ is enhanced by an order of magnitude and now exceeds
${C^P_2}^{(S)}$, and the tensor contribution is dominant for $l\le 4$. This is
contrary to the result in Figure~3 of \cite{cri93c} where the tensor
contribution is subdominant for all $l$. On the other hand, for a universe
with standard recombination, the tensor contribution is predominant for multiples with $l<30$ (\cite{cri93c}; \cite{ng96}). Thus, the tensor mode plays an important role in the large-scale polarization. Although the basis rotation effect raises considerably the low multipoles of the tensor spectrum, it only slightly alters the r.m.s. degrees of polarization that have been calculated in \cite{ng96}.

Now, we can construct the auto-correlation functions.
Spatial isotropy guarantees that
\begin{equation}
\langle t^*_{lm} t_{l'm'} \rangle =\frac{C_l}{2l+1}\delta_{ll'}\delta_{mm'}.
\label{tt}
\end{equation}
Hence, the temperature auto-correlation function is given by
$C(\Theta)\equiv \langle T_1 T_2 \rangle
= (1/4\pi)\sum_l C_l P_l(\cos\Theta)$,
where $\Theta$ is the separation angle. For the polarization, both $Q$ and $U$ only have cylindrical symmetry, so $q_{lm}$ and $u_{lm}$ do not
have a relation like equation~(\ref{tt}). As such, we first take a whole-sky average which is denoted by the curly brackets, and
then an average over all observational positions, namely,
\begin{equation}
C^P(\Theta)\equiv \langle\left\{Q_1 Q_2+U_1 U_2\right\}\rangle=
\frac{1}{4\pi}\sum_l C^P_l P_l(\cos\Theta),
\label{cf}
\end{equation}
where we have used $\left\{Y^*_{lm}({\bf \hat e_1})
Y_{l'm'}({\bf \hat e_2})\right\}=1/(4\pi)
P_l(\cos\Theta) \delta_{ll'}\delta_{mm'}$.
In actual observations, $Y_{lm}$ in equation~(\ref{tqu}) should be replaced by
$W_{lm}Y_{lm}$, where $W_{lm}$ is a window function appropriate to a particular experiment. However, for many polarization measurements the window function only depends on $l$ and is simply inserted in equation~(\ref{cf}).

\section{Cross-correlation Functions}

It is interesting and useful to know the cross correlations among $T$, $Q$,
and $U$ both theoretically and experimentally. The two-point cross-correlation functions $\langle T({\bf r},{\bf \hat e}) Q({\bf r},{\bf \hat z}) \rangle$,
$\langle T({\bf r},{\bf \hat e}) U({\bf r},{\bf \hat z}) \rangle$, and
$\langle Q({\bf r},{\bf \hat e}) U({\bf r},{\bf \hat z}) \rangle$ have been
considered, where the axes specifically chosen to define the Stokes parameters in the ${\bf\hat z}$ direction are $\bf\hat x$ and $\bf\hat y$ (\cite{cou94}; \cite{cri95}; \cite{kos96}). These cross-correlation functions with this particular choice of axes have limited use in the sense of actual observations. Essentially, this is because a single measurement of the Stokes parameters at two specific directions $\bf\hat e$ and $\bf\hat z$ has a large sample variance. However, this variance can be reduced by surveying a larger part of the sky, and it approaches the inevitable cosmic variance for a whole-sky coverage. As such, it is valuable to consider the whole-sky averaged
$\langle\left\{T_1 Q_2\right\}\rangle$,
$\langle\left\{T_1 U_2\right\}\rangle$, and
$\langle\left\{Q_1 U_2\right\}\rangle$.

Let us consider $\langle\left\{T_1 U_2\right\}\rangle$ first, from
equations~(\ref{tqu}) and (\ref{ps}), it is given by
\begin{eqnarray}
\langle\left\{T_1 U_2\right\}\rangle &=& \frac{1}{4\pi}\sum_l
\langle\sum_m t^*_{lm} u_{lm} \rangle P_l(\cos\Theta), \nonumber \\
\langle\sum_m t^*_{lm} u_{lm} \rangle &=& \frac{2l+1}{4\pi}
\int d{\bf\hat e_1}d{\bf\hat e_2} P_l({\bf \hat e_1\cdot \hat e_2})
\langle T_1 U_2 \rangle.
\end{eqnarray}
For the scalar mode, we obtain
\begin{eqnarray}
\langle\sum_m t^*_{lm} u_{lm} \rangle&=&-\frac{2l+1}{4\pi}
\sum_{l_1 l_2} (2l_1+1)(2l_2+1)
\int k^2 dk\; \alpha_{l_1} \beta_{l_2}
\int d{\bf\hat e_1}d{\bf\hat e_2}d{\bf\hat k} \nonumber \\
&&P_l({\bf \hat e_1\cdot \hat e_2}) P_{l_1}({\bf \hat e_1\cdot \hat k})
P_{l_2}({\bf \hat e_2\cdot \hat k}) \sin 2\psi_2.
\label{tu}
\end{eqnarray}
Using the properties that $\sin 2\psi$ is odd under the parity transformation:
${\bf\hat e}\to{-\bf\hat e}$, and even under the transformation:
${\bf\hat k}\to{-\bf\hat k}$, one can easily show that
$\langle\sum_m t^*_{lm} u_{lm} \rangle$ in equation~(\ref{tu}) is identically
zero, hence, $\langle\left\{T_1 U_2\right\}\rangle=0$. Similar
argument shows that $\langle\left\{Q_1 U_2\right\}\rangle=0$, but it does not guarantee that $\langle\left\{T_1 Q_2\right\}\rangle$ vanishes. However, 
by direct evaluating the multi-dimensional integral, we have found that $\langle\left\{T_1 Q_2\right\}\rangle=0$. For the tensor-mode case,
similarly, using the parity properties of the angles $\theta$, $\phi$, and $\psi$, one can show that $\langle\left\{T_1 U_2\right\}\rangle=0$, and
$\langle\left\{Q_1 U_2\right\}\rangle=0$. Notice that the interference terms between the cross and plus modes in the integrals vanish as expected.
For $\langle\left\{T_1 Q_2\right\}\rangle$, we have evaluated the integral
numerically and the result is consistent with zero.

As a result, all full-sky averaged two-point cross-correlation functions vanish. This does not mean that there is no correlation between anisotropy and polarization. In fact, the correlated part has been extracted, and its polarization pattern has been correlated with the cold and hot spots of the microwave sky (\cite{cou94}; \cite{cri95}). The zero result here only suggests that the correlated part has inherent symmetry over the whole sky.  It is interesting to estimate how much the correlation is suppressed by averaging over an area of sky big enough to reduce the sample variance to acceptable levels. However, it can hardly be done without knowing the inherent symmetry, and the result should depend on sky sampling. Therefore, it is important to find out the exact (no approximations or whole-sky averaging) correlation functions of the Stokes parameters between any two arbitary points on the celestial sphere. This would reveal the symmetry, and be useful to measurements.

We have tried to calculate a special case that is 
$\langle T({\bf r},{\bf \hat z}) Q({\bf r},{\bf \hat e}) \rangle$.
For the scalar case, the result is 
\begin{eqnarray}
&&\langle T({\bf r},{\bf \hat z}) Q({\bf r},{\bf \hat e}) \rangle
=2\pi\sum_{l\ge 2} (2l+1) C_l^{TQ} P_l^2(\cos\theta), \nonumber \\
&& C_l^{TQ} = \frac{(l-2)!}{(l+2)!} \sum_{l'}(2l'+1)
\left[\int_{-1}^1 dx P_l^2(x) P_{l'}(x) \right]
\int k^2 dk \alpha_l^* \beta_{l'},
\label{tq}
\end{eqnarray}
which is similar to the result given in \cite{cou94} except that our result has no azimuthal dependence. This is expected because $Q$ has a cylindrical symmetry. Equation~(\ref{tq}) would be useful to the on-going polarization measurement that would observe an annulus of $10$ spots at constant declination (P. T. Timbie 1995, private communication). If the spots are well separated, then the measured anisotropy-polarization correlation will be
\begin{equation}
T_0 \bar Q =\langle TQ(\theta)\rangle \pm\frac{\sigma_T\sigma_Q}{\sqrt n},
\end{equation}
where $\bar Q=\sum_{i=1}^n Q_i/n$, $\sigma_T$ and $\sigma_Q$ are the anisotropy and polarization variances respectively, $T_0$ is the measured polar temperature anisotropy and $n=10$. (We have neglected the detector noise.) In a reionized universe, $\langle TQ(\theta=50^o)\rangle\simeq -0.01\sigma_T\sigma_Q$ (\cite{cou94}), hence the expected signal to noise ratio $S/N\simeq 0.03$. When $\theta=4^o$, $S/N\simeq 0.3$. Whereas, the $S/N$ ratio of the polarization variance $\langle Q^2 \rangle$ is $\sqrt 5$. Note that one is likely to find a relatively large $\bar Q$ when $T_0$ is not at a hot or cold spot on the microwave sky.

\section{Conclusions}

We have calculated all whole-sky averaged two-point correlation functions of
the Stokes parameters of the CMB. Incorporating the basis rotation effects, we have found that the tensor contribution to the polarization power spectrum get enhanced for low multipoles whereas the scalar contribution remains unchanged, and that all full-sky averaged cross-correlation functions vanish.
A special case of $\langle TQ\rangle$ has been calculated, and used to show that the signal to noise of $\langle TQ\rangle$ is low for the polarization measurement underway.  We intend to calculate the exact correlation functions that would be important to future polarization experiments.

\acknowledgments

This work was supported in part by the R.O.C. NSC grant NSC85-2112-M-001-014.

\clearpage

\figcaption[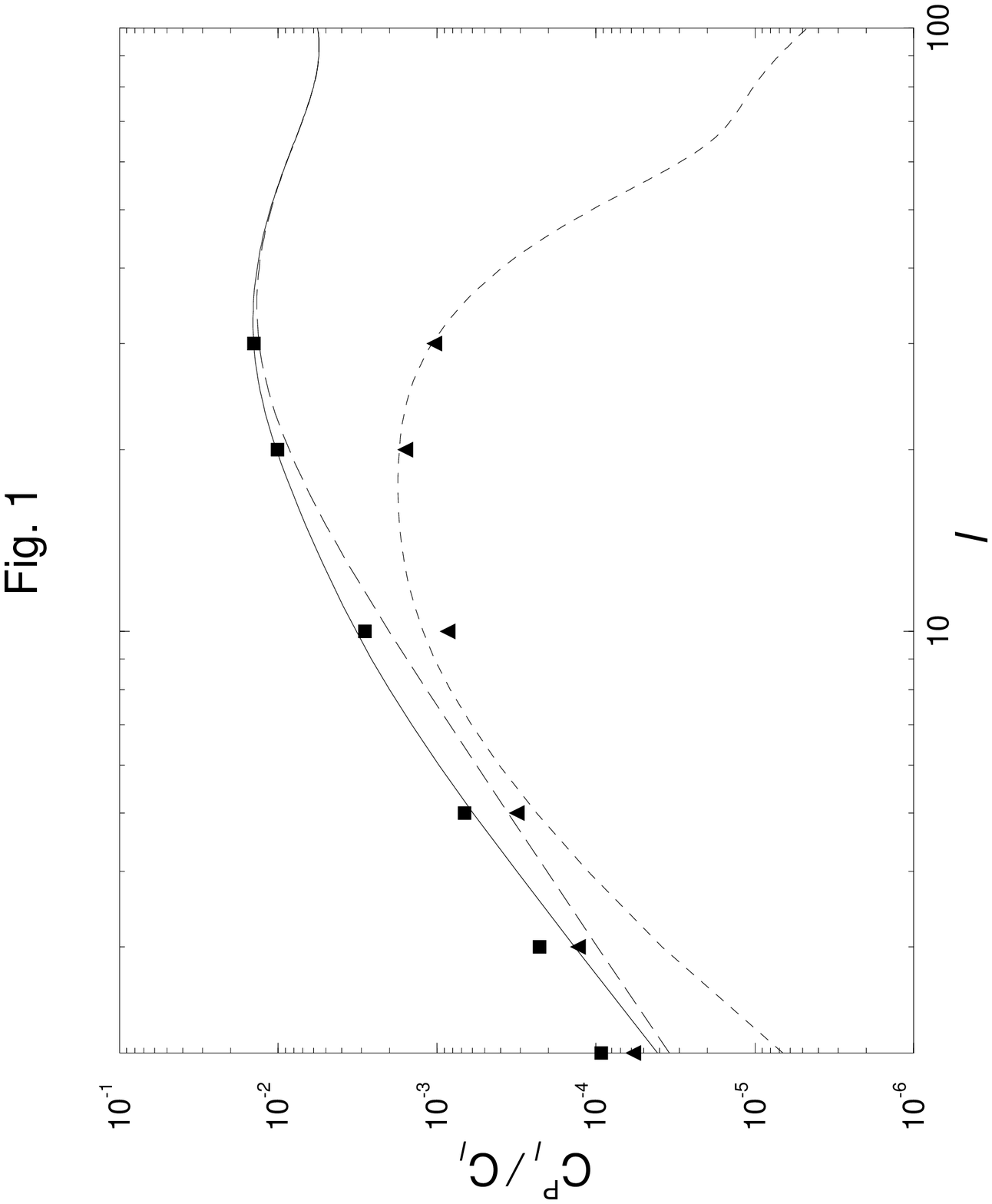]{Ratio of polarization multipole to anisotropy multipole vs. $l$ in a standard CDM model with no recombination. The short-dashed, long-dashed, and solid curves denote respectively the tensor, scalar, and total contributions, using
the approximate formulae with no rotation angles. The solid triangles and
squares denote respectively the corrected tensor and total contributions.
\label{fig1}}

\clearpage

\plotone{fig1.eps}


\clearpage


\begin{thebibliography}{}

\bibitem[Bond \& Efstathiou 1987]{bon87} Bond, J. R., \& Efstathiou, G. 1987,
         \mnras, 226, 655

\bibitem[Coulson, Crittenden, \& Turok 1994]{cou94} Coulson, D.,
         Crittenden, R. G., \& Turok, N. G. 1994, Phys. Rev. Lett., 73, 2390

\bibitem[Crittenden, Coulson, \& Turok 1995]{cri95} Crittenden, R. G.,
         Coulson, D., \& Turok, N. G. 1995, Phys. Rev. D52, R5402

\bibitem[Crittenden, Davis, \& Steinhardt 1993]{cri93c}
         Crittenden, R., Davis, R. L., \& Steinhardt P. J. 1993, ApJ, 417, L13

\bibitem[Crittenden et al. 1993]{cri93a} Crittenden, R., et al. 1993, Phys.
         Rev. Lett., 71, 324

\bibitem[Kosowsky 1996]{kos96} Kosowsky, A. 1996, Ann. Phys., 246, 49

\bibitem[Ng \& Ng 1995]{ng95} Ng, K. L., \& Ng, K.-W. 1995, \apj, 445, 521

\bibitem[Ng \& Ng 1996]{ng96} Ng, K. L., \& Ng, K.-W. 1996, \apj, 456, 413

\bibitem[Lubin, Melese, \& Smoot 1983]{lub83} Lubin, P., Melese, P., \& Smoot,          G. 1983, \apj, 273, L51

\bibitem[Partridge, Nowakowski, \& Martin 1988]{par88} Partridge, R. B.,          Nowakowksi, J., \& Martin, H. M. 1988, Nature (London), 331, 146
    
\bibitem[Smoot et al. 1992]{smo92} Smoot, G. F., et al. 1992, \apjl, 396, L1

\bibitem[Steinhardt 1994]{ste94} Steinhardt, P. J. 1994, Proceedings of the
         1994 Snowmass Summer Study, E. W. Kolb and R. D. Peccei, World
         Scientific, 1995, 51

\bibitem[Wollack et al. 1993]{wol93} Wollack, E. J., et al. 1993, \apjl, 419,
         L49

\end{thebibliography}
\end{document}